\documentclass[aps,prl,twocolumn,amsmath,amssymb]{revtex4}
\usepackage{graphicx,epsfig,epsf}
\usepackage{dcolumn}
\usepackage{bm,psfrag,color}
\usepackage[matrix,frame,arrow]{xypic}

\begin{document}

\newcommand{\ri}{{\rm i}}
\newcommand{\re}{{\rm e}}
\newcommand{\bx}{{\bf x}}
\newcommand{\bzero}{{\bf 0}}
\newcommand{\bd}{{\bf d}}
\newcommand{\br}{{\bf r}}
\newcommand{\bk}{{\bf k}}
\newcommand{\bA}{{\bf A}}
\newcommand{\bB}{{\bf B}}
\newcommand{\bb}{{\bf b}}
\newcommand{\bD}{{\bf D}}
\newcommand{\bE}{{\bf E}}
\newcommand{\bj}{{\bf j}}
\newcommand{\bI}{{\bf I}}
\newcommand{\bH}{{\bf H}}
\newcommand{\bJ}{{\bf J}}
\newcommand{\bR}{{\bf R}}
\newcommand{\bM}{{\bf M}}
\newcommand{\bem}{{\bf m}}
\newcommand{\bn}{{\bf n}}
\newcommand{\bs}{{\bf s}}
\newcommand{\bp}{{\bf p}}
\newcommand{\tr}{{\rm tr}}
\newcommand{\tbs}{\tilde{\bf s}}
\newcommand{\rSi}{{\rm Si}}
\newcommand{\beps}{\mbox{\boldmath{$\epsilon$}}}
\newcommand{\bsigma}{\mbox{\boldmath{$\sigma$}}}
\newcommand{\bmu}{\mbox{\boldmath{$\mu$}}}
\newcommand{\bthe}{\mbox{\boldmath{$\theta$}}}
\newcommand{\rg}{{\rm g}}
\newcommand{\xmax}{x_{\rm max}}
\newcommand{\ra}{{\rm a}}
\newcommand{\rx}{{\rm x}}
\newcommand{\rs}{{\rm s}}
\newcommand{\rP}{{\rm P}}
\newcommand{\up}{\uparrow}
\newcommand{\down}{\downarrow}
\newcommand{\hc}{H_{\rm cond}}
\newcommand{\kb}{k_{\rm B}}
\newcommand{\cI}{{\cal I}}
\newcommand{\tit}{\tilde{t}}
\newcommand{\cE}{{\cal E}}
\newcommand{\cC}{{\cal C}}
\newcommand{\Ubs}{U_{\rm BS}}
\newcommand{\qq}{{\bf ???}}
\newcommand{\pob}{\overline{\phi}_0}
\newcommand*{\etal}{\textit{et al.}}

\sloppy

\title{A simpel and versatile cold-atom simulator of 
  Non-Abelian gauge potentials}
\author{Daniel Braun$^{1,2}$ }

\affiliation{$^{1}$Universit\'e de Toulouse, UPS, Laboratoire
de Physique Th\'eorique (IRSAMC), F-31062 Toulouse, France}
\affiliation{$^{2}$CNRS, LPT (IRSAMC), F-31062 Toulouse, France}
\begin{abstract}
We show how a single, harmonically trapped atom in a
  tailored magnetic field can be used for simulating the effects of a
  broad class of non-abelian gauge potentials. We
  demonstrate how to implement Rashba or Linear-Dresselhaus couplings,
  or observe {\em Zitterbewegung} of a Dirac particle.  
\end{abstract}

\maketitle

Simulating complex quantum systems with the help of trapped cold atoms
has become a flourishing branch of quantum optics. The exquisite
control and complex internal structure of cold atoms
have allowed successful experimental simulation of systems and effects ranging
from quantum phase transitions in the
Bose-Hubbard model \cite{Greiner02}, over Anderson localization
\cite{Billy08,Roati08}, to cosmological models \cite{Leonhardt00}.
Recently, a lot 
of research efforts 
have been directed towards realization of Abelian or Non-Abelian gauge
fields.  By bathing an optical lattice in additional weak,
non-resonant light that can create Raman-transitions between
hyperfine levels, one can create artificial magnetic fields which can
be extremely strong. Vortex formation in a BEC has been observed due to such
artificial magnetic fields \cite{Lin09}.  Even the creation of magnetic
monopoles  has been proposed \cite{Pietilae09}. Non-abelian gauge fields can
be useful 
for studying spin-tronics materials with various spin-orbit couplings,
Berry phases, or topologically protected qubits.

Most of these quantum simulations have been proposed or performed for
atoms trapped in an optical lattice or Bose-Einstein condensates,
requiring rather sophisticated experimental setups. Many 
physical systems are, however, interesting as single-particle systems.
Such is the case e.g.~for Anderson localization, the low-energy behavior of
electrons in Graphene, or the relativistic motion of electrons that
leads to the effect of {\em Zitterbewegung} \cite{Schroedinger30}.  It would 
be highly desirable to have a simple, versatile 
system that allows to simulate such single particle dynamics. We show here
that such a system can 
be constructed from a single, harmonically trapped atom exposed to a
suitably tailored real (physical) magnetic field.  We show that by
simply changing some gradients of different field components different
non-Abelian gauge fields can be simulated, giving rise, for example, to Rashba
or linear Dresselhaus coupling.  We also show that {\em Zitterbewegung}
should be easily observable.  Abelian gauge fields can also be obtained, but
are  less interesting in the proposed setup, as they do not depend on
position.

Consider a single atom of mass $m$ trapped in a harmonic potential
(frequency $\omega$) and exposed to
a magnetic field $\bB(\bx,t)$. For concreteness, let us assume a
neutral atom
in an optical dipole trap.  The magnetic field should not be used 
for any part of the trapping, but 
be controllable independently of the trapping. We will also assume that
the atom is  
cooled close to the ground state such that approximating the trap by a
harmonic potential in all three directions is indeed reasonable. The
hamiltonian of the system reads 
\begin{equation} \label{H}
H=\frac{\bp^2}{2m}+\frac{1}{2}m\omega^2\bx^2+g_J\mu_B\bB(\bx,t)\cdot\bJ/\hbar\,, 
\end{equation}
where $g_J$, $\mu_B$, and $\bJ$ are the g-factor, Bohr-magneton, and total
angular momentum of the atom, and we have neglected the nuclear spin
contribution in the Zeeman term. Suppose that the magnetic field varies slowly
on the scale of the trapping potential, such that we can expand it in a
power series about the origin $\bx=0$, 
\begin{equation} \label{Bex}
\bB(\bx,t)=\bB(\bzero,t)+\sum_{i=1}^3\nabla B_i\cdot\bx\,\hat{e}_i+{\cal
  O}(\bx^2)\,,
\end{equation}
with unit vectors $\hat{e}_i$ in directions $i=1,2,3\equiv x,y,z$.
We have neglected higher order terms starting with the second order. One can
still include the second
order, with consequences to be discussed below, but for the moment suppose
that the magnetic field varies slowly enough over the length-scale of the
atomic motion in the trap that the quadratic term in $\bx$ from expanding
$\bB$ can indeed be neglected compared to the quadratic term describing the
trapping potential. 

The linear term in $\bx$ then leads to a coupling
$\bx\cdot\nabla(\bB\cdot\bJ)\equiv\bx\cdot\sum_i(\nabla
B_i\left.\right|_{\bx=\bzero})J_i$.
Now let us 
canonically transform $\bx/x_0\rightarrow -\bp/p_0$, $\bp/p_0\rightarrow
\bx/x_0$, where I have introduced the natural scales $x_0$ and $p_0$ of the
canonical coordinates and momenta of the atom in the trap,
$x_0=\sqrt{\hbar/(m\omega)}$ and $p_0=\sqrt{\hbar m\omega}$.  The
harmonic oscillator 
part in the hamiltonian is invariant under this transformation, but the
$\bx$ in the Zeeman term becomes a $-\bp$, and we get the transformed
hamiltonian $\tilde{H}$,
\begin{equation} \label{Ht}
\tilde{H}
=\frac{\bp^2}{2m}+\frac{1}{2}m\omega^2\bx^2-\frac{e}{m}\bp\cdot\bA+g_J\mu_B\bB(\bzero,t)\cdot\bJ/\hbar\,,  
\end{equation}
with 
\begin{equation} \label{A}
\bA=\frac{\mu_Bg_J}{e\hbar\omega}\nabla\left(\bB\cdot\bJ\right)_{\bx=\bzero}\,.
\end{equation}
Note that $\bA$ is independent of position and thus commutes with $\bp$. We
can therefore rewrite the hamiltonian in the form
\begin{equation} \label{Ht2}
\tilde{H}=\frac{1}{2m}(\bp-e\bA)^2+\frac{1}{2}m\omega^2\bx^2+\phi+\mu_Bg_J\bB(\bzero,t)\cdot\bJ/\hbar\,,
\end{equation}
which makes appear $\bA$ as a gauge potential. The constant $\phi$ acts only
in the angular--momentum Hilbert space, 
\begin{equation} \label{phi}
\phi\equiv -\frac{e^2\bA^2}{2m}\,,
\end{equation}
and can be considered an anisotropy for the angular momentum. We can
write it as 
\begin{equation} \label{phi2}
\phi=-\frac{1}{2}\bJ^t I^{-1}\bJ\,,
\end{equation}
where the inverse tensor of inertia $I^{-1}$ has matrix elements
\begin{equation} \label{I-1}
(I^{-1})_{ij}=\frac{\mu_B^2g_J^2}{m\hbar^2\omega^2}\nabla B_i\cdot\nabla B_j\,.\end{equation}
The effect created by the gauge potential plays an appreciable
role only if the components of $e\bA$ are comparable to the corresponding
components of $\bp$. The latter are, close to the ground-state of the
harmonic potential, of the order of $p_0$. With
$\bJ\sim\hbar$, we are thus 
led to a condition for the magnitude of the magnetic field gradient $B'$
\begin{equation} \label{cond}
B'\sim\frac{\sqrt{\hbar m}\omega^{3/2}}{\mu_B}\,,
\end{equation}
which should be satisfied for at least one component of $\bB$. Inserting the
data for $^{87}$Rb, we find a gradient $B'=2.1$Gauss/mm for
$\omega=2\pi\cdot$1kHz, which appears to be a very convenient order of
magnitude. \\

It is worthwhile spelling out the components of $\bA$ explicitly. Up
to the common prefactor $\frac{\mu_Bg_J}{e\hbar\omega}$ we have
\begin{eqnarray} \label{Acom}
\bA_x&\propto&\partial_xB_x\,J_x+\partial_xB_y\,J_y+\partial_xB_z\,J_z\\
\bA_y&\propto&\partial_yB_x\,J_x+\partial_yB_y\,J_y+\partial_yB_z\,J_z\\
\bA_z&\propto&\partial_zB_x\,J_x+\partial_zB_y\,J_y+\partial_zB_z\,J_z\,.
\end{eqnarray}
The components of $\bJ$ are real physical angular momentum, and
thus satisfy $[J_x,J_y]=i\hbar J_z$ in any Hilbert space.  Therefore,
obviously, the gauge field is in general non-abelian.  The
remarkable thing about $\bA$ is that it is easily programmable by an
appropriate choice of magnetic field gradients.  Choosing for instance only
a field 
gradient of one magnetic field component leaves us with only one component
of the angular 
momentum and we thus get an abelian gauge potential. Choosing arbitrary
linear combinations of angular momentum operators is only restricted by the
properties of the real physical magnetic field employed, in particular
${\rm div}\bB=0$, as we shall see below.

Before working out a few specific examples, three more remarks:
\begin{enumerate}
  \item We chose the same trap frequency $\omega$ in all directions,
 which is certainly somewhat unusual for an optical dipole trap.  After
 canonical 
 transformation, this translates into the same mass for all three
 directions.  Had we different trap frequencies in different directions, we
 would get effectively different masses in different directions, i.e.~a
 (diagonal) effective mass tensor, similar to the situation in a
 semiconductor. This may or may not be useful, depending on what one wants
 to simulate.
\item Similarly, if we kept the quadratic term in the expansion of $\bB$,
  we would get additional quadratic terms (which would also depend on the
  internal state of the atom) contributing to the trapping potential, such
  that, again, after the canonical transformation we would end up with
  different 
  masses in different directions.
\item What was derived here for a single atom should translate
  immediately to a BEC, if the interactions between atoms are
  negligible. If they are not, any interaction potential $V(\bx_i-\bx_j)$
  will translate into a momentum dependent interaction after the canonical
  transformation. 
\end{enumerate}
Let us now look at some concrete examples of non-Abelian gauge
potentials, inspired by the list in \cite{Vaishnav08}. I will assume
that we have an atom 
with $J=1/2$, so that the $J_i$ 
are given by Pauli-matrices, $J_i=(\hbar/2)\sigma_i$.\\

{\em Rashba coupling.}
This is a coupling $\alpha(p_y\sigma_z-p_z\sigma_y)$, with some constant
$\alpha$. It can easily be achieved by choosing $\partial_xB_i=0$, for
$i=x,y,z$, $\partial_y 
B_y=0=\partial_z B_z$, and $\partial_y B_z=B'=\partial_z B_y$.  No magnetic
mono-pole is required, ${\rm div}\bB=0$, but the field has a finite curl in
$x$-direction, $(\nabla\times\bB)_x=2B'$. That curl is needed 
at $\bx=\bzero$, where we did the expansion of the magnetic field. But
since we insisted on a linearly growing field over the lengthscale of the
motion in the trap, we request basically circular magnetic field lines around
the $x$-axis, with magnetic field
strength growing linearly with distance from the $x$-axis.  That is, of
course, quite different from what a current carrying wire along the
$x-$axis would do (apart from the fact 
that that wire would have to go straight through $\bx=0$, where we
would like 
to trap the atom).  A possible way of creating the required magnetic field
might be
through an electric field that increases linearly in time by using segmented
cylindrically arranged electrodes. Choosing the right voltage profile for
all the electrodes one can create a radially symmetric electric
field 
oriented in $x$-direction that increases proportionally in radial direction,
and generates the desired $\bB$-field according to Maxwell's equation
$\nabla\times\bH=\dot{\bD}+\bj$. 

The inverse tensor of inertia is here simply $I^{-1}={\rm
  diag}(0,B'^2,B'^2)$ and just leads to a constant $\phi=B'^2{\bf 1}$,
  i.e.~an identity matrix in spin Hilbert--space. \\

{\em Linear Dresselhaus coupling.} 
Here we want $A_y=\alpha\sigma_y$, $A_z=-\alpha\sigma_z$. Choose again
$\partial_x B_i=0$ $\forall i$, but then $\partial_y B_y=B'=-\partial_z
B_z$, and all other derivatives equal zero. Again, this is compatible with
${\rm div} \bB=0$, and we basically get a field that is point symmetric,
growing 
linearly in radial direction, but pointing towards the center on the
$z-$axis and outward on the $y-$axis, like from a quadrupole magnet.

Inverse tensor of inertia and the resulting potential $\phi$ are identical
to the Rashba case.\\

{\em Graphene sheet in vicinity of Dirac point.}
This case is more problematic: we want $A_y=\alpha\sigma_y$,
$A_z=\alpha\sigma_z$, which leads to $\partial_y B_y=B'=\partial_z B_z$. But
${\rm div} \bB=0$ then requires $\partial_xB_x=-2B'$, and thus adds an extra
coupling term $-2p_x\sigma_x$ to the desired $p_y\sigma_y+p_z\sigma_z$. \\

{\em Zitterbewegung.}
{\em Zitterbewegung} (ZB) is an interference effect first predicted by
Schr\"odinger for 
relativistic spin-1/2 particles that leads to a jittering motion on
the length scale of the Compton wavelength of the particle,
i.e.~$h/mc\simeq 10^{-12}$m for an electron \cite{Schroedinger30}.
This short length scale 
has so far prevented direct experimental observation of ZB for
relativistic electrons.  However, the effect should exist for any
spinor system with linear dispersion relation.  Consequently, ZB has
been studied theoretically in several systems, including mesoscopic
wires \cite{Schliemann05}, graphene \cite{Cserti06}, ion traps
\cite{Lamata07}, and optical lattices \cite{Vaishnav08}. Rashba
couplings in quantum dots with rotationally invariant potentials in 
2D was studied in \cite{Tsitsishvili04}.  Very
recently, ZB was observed experimentally with a single trapped ion
\cite{Gerritsma09}. 
ZB has mostly been studied for a free particle. The ZB then dies out
after a short time, 
when the two wave-packages corresonding to spin-up and spin-down have
se\-pa\-ra\-ted enough to prevent further interference.  It is
interesting to consider how the additional harmonic confinement
potential in (\ref{Ht2}) modifies the ZB. One might expect that the
confinement will increase the time interval in which the ZB can be observed. \\

We now study ZB for the harmonic confinement (\ref{Ht2}) in the
Rashba case. We show that ZB exists even if the harmonic oscillator is
initially unexcited, 
and that it can persist for arbitrarily long times.
Note that the 
third direction, $x$ in the above notation, is not affected by the gauge
field and just se\-pa\-ra\-tes. It is convenient then, to rewrite
(\ref{Ht2}) in terms of annihilation (creation) operators
$a(a^\dagger)$ and $b(b^\dagger)$ for the $y$ and $z$ components,
respectively. We are thus lead to 
\begin{eqnarray}
H&=&\hbar\omega\left(a^\dagger a+b^\dagger b +i b_1
\big(-(a-a^\dagger)\sigma_z+(b-b^\dagger)\sigma_y\right)\nonumber\\
&&+\bb_0\cdot\bsigma\big)\,, 
\end{eqnarray}
where the dimensionless parameters are
$b_1=\frac{g_J\mu_B B'}{2\sqrt{\hbar m \omega^3}}$ and
$\bb_0=\frac{\mu_Bg_J\bB(0)}{2\hbar\omega}$, and we have suppressed an
irrelevant constant. 

It is straightforward to express $H$ in basis
states $|nms\rangle$, where $n,m\in\{0,1,2,\ldots\}$ are the occupation
  numbers for oscillators $a$ and $b$, and $s\in\{\pm 1\}$ label
  eigenstates of $\sigma_z$ corresponding to these eigenvalues, and to
  diagonalize the Hamiltonian numerically.  This basis turns out to be highly
  suitable --- taking into account only up to 5 excitations per
  oscillator (i.e.~a 72 dimensional basis) allows one to find the lowest
  26 eigenstates already with 4 significant digits for $b_1=0.1$
  (compared to the 
  case with 10 excitations per oscillator, or 242 basis states, which
  we used in the numerical simulations).  With
  the obtained propagator we can study 
  the time evolution of the averages of the observables $\langle
  y(t)\rangle$, $\langle p_y(t)\rangle$ etc. 

\begin{figure}
\epsfig{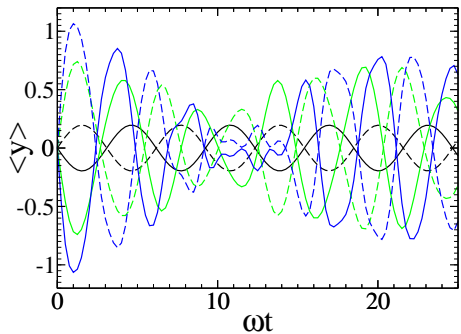}
\hspace{0.3cm}
  \caption{(Color online) {\em Zitterbewegung} in $\langle y(t)\rangle$
  (in units of 
  harmonic oscillator length $x_0$) for initial states $|00+\rangle$
  (full lines) and $|00-\rangle$ (dashed lines); $\bb_0=\bzero$, and $b_1$
  takes the values $b_1=0.1$, $0.5$ and 1 for black lines, green
  (light grey)
  lines, and blue lines (dark grey), respectively. 
  }\label{fig.ZB}
\end{figure}

Figure \ref{fig.ZB} shows
  $\langle y(t)\rangle$ for $\bb_0=\bzero$, different values of
  $b_1$, and the two initial states $|\psi(0)\rangle=|00+\rangle$ and
  $|00-\rangle$. Both $\langle z(t)\rangle$ and $\langle
  p_z(t)\rangle$ remain always zero in these cases, whereas $\langle
  y(t)\rangle$ shows oscillations. They are always with opposite sign
  for these two different initial states. For small values of $b_1$
  ($b_1=0.1$) the oscillations appear periodic, whereas with
  increasing $b_1$, additional harmonics appear that make the signal
  look more and more erratic. By changing the initial state of the
  spin, the direction of the oscillation can be chosen.  
  E.g.~for$|\psi(0)\rangle=(|00+\rangle+i|00-\rangle)/\sqrt{2} $, we
  obtain  $\langle y(t)\rangle=0=\langle p_y(t)\rangle$, whereas the
  $z$-component now shows the signal we had for $\langle
  y(t)\rangle$. The state
  $|\psi(0)\rangle=(\cos(\pi/8)|00+\rangle+i\sin(\pi/8)|00-\rangle)$
  leads to oscillations with $\langle y(t)\rangle=-\langle
  z(t)\rangle$. It appears thus that the ZBis always
  one-dimensional as long as the initial state is chosen in the
  subspace $\{|00+\rangle,|00-\rangle\}$. The superposition
  $|\psi(0)\rangle=(|00+\rangle+|00-\rangle)/\sqrt{2} $ switches the
  ZB
  off in both components. Note that one-dimensional ZB was also predicted for
  a 1D harmonic confinement and a specific initial spin state, with
  the ZB perpendicular to the 
  free 1D motion \cite{Schliemann05}.\\

\begin{figure}
\epsfig{file=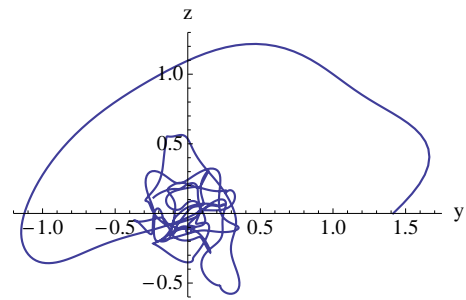,width=6cm,angle=0}
\hspace{0.3cm}
  \caption{(Color online) Trajectory of 2D {\em Zitterbewegung} in $(\langle
  y(t)\rangle,\langle z(t)\rangle)$ 
  (in units of 
  harmonic oscillator length $x_0$) for an initial state
  $|\alpha\beta+\rangle$, where $|\alpha\rangle$ and $|\beta\rangle$ are
  coherent states with $\alpha=1$ and $\beta=i$ at $\bb_0=\bzero$ and
  $b_1=1$. The trajectory is shown for $0\le \omega t\le 50$. The initial 
  state is localized at $\langle y(0)\rangle=\sqrt{2}=\langle p_z(0)\rangle$,
  $\langle z(0)\rangle=0=\langle p_y(0)\rangle$ and would lead for $b_1=0$ to circular 
  motion in the $yz$-plane. 
  }\label{fig.ZB2D}
\end{figure}

\begin{figure}
\epsfig{file=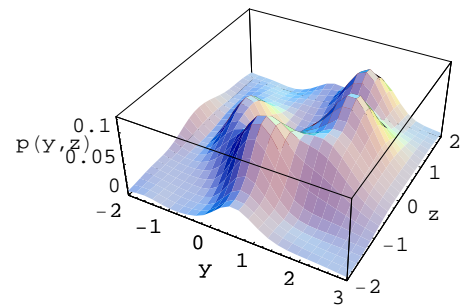,width=6cm,angle=0}
\hspace{0.3cm}
  \caption{(Color online) Probability distribution $p(y,z)$ of finding the
    atom localized at position $y,z$ (in units of $x_0$) at $\omega t=10$,
    for $b_1=1.0$, $\bb=\bzero$. The initial coherent state (same
state as in Fig.\ref{fig.ZB2D}) gets smeared out due to the anharmonicity
created by the Rashba coupling. }\label{fig.ZB2Dp}
\end{figure}

The results for small $b_1$ can be easily understood analytically, by going
to the Heisenberg picture and expanding the time dependent operators to
lowest order in $b_1$.  We find, correct to order ${\cal O}(b_1)$, at
$\bb_0=\bzero$ 
\begin{eqnarray}
\langle y(t)\rangle&\simeq&\langle y(0)\rangle\cos\omega t +\langle
p_y(0)\rangle\sin\omega t \nonumber\\
&&+\sqrt{2}b_1
\langle \sigma_z(0)\rangle\sin\omega t\label{yt}\\
\langle z(t)\rangle&\simeq&\langle z(0)\rangle\cos\omega t
+\langle p_z(0)\rangle\sin\omega t \nonumber\\
&&-\sqrt{2}b_1\langle \sigma_y(0)\rangle\sin\omega t \\
\langle p_y(t)\rangle&\simeq&\langle p_y(0)\rangle\cos\omega t -\langle
y(0)\rangle\sin\omega t \nonumber\\
&&+\sqrt{2}b_1\langle \sigma_z(0)\rangle(\cos\omega t-1)\\
\langle p_z(t)\rangle&\simeq&\langle z(0)\rangle\cos\omega t
-\langle z(0)\rangle\sin\omega t \nonumber\\
&&-\sqrt{2}b_1\langle \sigma_y(0)\rangle(\cos\omega t-1)\label{pzt} \,.
\end{eqnarray}
All positions and momenta are expressed in terms of the length-scales $x_0$
and momentum scales $p_0$ of the harmonic oscillator, respectively.  The
corrections to Eqs.(\ref{yt}-\ref{pzt}) are of order ${\cal O}(B_1^2)$. We
recognize in the terms independent of $b_1$ the unperturbed 
motion of the 2D harmonic oscillator.  This motion can be switched off by
chosing average initial positions and momenta equal zero, as is the case for
the initial states discussed above with the two harmonic oscillators
initially in their ground state.  All motion is then entirely due
to the ZB and indeed 1D, in a direction given by the vector
$(\langle\sigma_z(0)\rangle, \langle\sigma_y(0)\rangle)$.
Interestingly, at $|b_1|\ll 1$ the ZB is itself harmonic with an amplitude 
controlled by $b_1$. The fact that the length and momenta scales of the ZB
are set by the harmonic confinement should allow for a much simpler
experimental verification than for Dirac electrons.    Moreover, the
harmonic confinement potential keeps the wavepackages corresponding to
the two different spin components together, preventing the decay of
the ZB, which should facilitate its experimental study. A damping of the ZB
can arise at higher values of $b_1$ and initial coherent states of
the harmonic oscillators, which will in general be smeared out due to the
anharmonicity mediated by the coupling to the spin (see Figs.\ref{fig.ZB2D}
and \ref{fig.ZB2Dp}). At $b=1$ the dynamics looks random and diffusive
in the $yz$-plane, but the trajectory of average values $(\langle
y(t)\rangle, \langle  z(t)\rangle)$ is of course entirely deterministic
and reproducible.

\begin{figure}
\epsfig{file=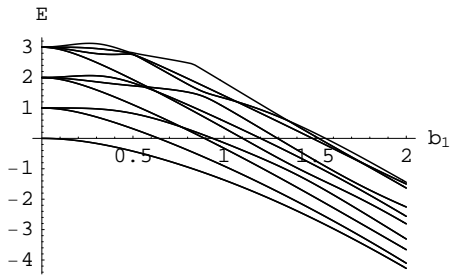,width=6cm,angle=0}\hspace{0.3cm}
  \caption{Lowest $20$ energies in the Spectrum of $H$
  in units of $\hbar\omega$ as 
  function of $b_1$ at $\bb_0=\bzero$. The spectrum is symmetric under
  $b_1\to-b_1$, and each energy level shown is still doubly degenerate.
  }\label{fig.spec}
\end{figure}

The increasingly erratic behavior of the ZB with increasing $b_1$ can
be understood by looking at the spectrum of $H$, plotted in
Fig.\ref{fig.spec}. The oscillator states at $b_1=0$ containing $n$
energy quanta which are $2(n+1)$--fold degenerate (states
$|n0+\rangle$, $|n-1\,1+\rangle,\ldots, |0n+\rangle$, and the same set
once more but with spin down). They split with increasing $b_1$ and lead to
several avoided crossings. The resulting incommensurate frequencies
lead to the observed quasi-periodic, apparently random behavior.    

{\em In summary,} we have shown how a single trapped atom with hyperfine
structure trapped in a magnetic field with suitably tailored field gradients
can be used to simulate the effect of non-abelian gauge potentials.  We have
demonstrated how different effective spin-orbit couplings (such as Rashba or
linear Dresselhaus couplings) can be easily obtained, and we have proposed
a new way of observing {\em  Zitterbewegung} of a harmonically trapped
particle. An immediate consequence of the fact that (\ref{Ht2}) was obtained by
a canonical transformation that exchanges position and momentum, is
that the ZB in $y(t)$ will, of course, show up in $p_y(t)$ in the
original system. 
The simplicity and flexibility of the proposed setup
may also allow the study of other spin-orbit couplings, as well as
of applications such as robust quantum gates based on topological
phases \cite{Kitaev95}. 


{\em Acknowledgements:} I would like to thank the JQI (University of Maryland
and NIST Gaithersburg) for hospitality
during my stay during which this work was started, and Victor Galitski, Trey
Porto, Bill Phillips, Ian Spielman, and Jay Vaishnav for
discussions.

\bibliography{../mybibs_bt}

\end{document}